\documentclass{aastex}
\usepackage{emulateapj5,psfig}
\usepackage{graphicx}
\usepackage{times}
\makeatletter
{\end{center}\end{minipage}\smallskip}

\newenvironment{inlinefigure}{%
\def\@captype{figure}%
\noindent\begin{minipage}{0.999\linewidth}\begin{center}}
{\end{center}\end{minipage}\smallskip}
\makeatother
\def\ltsima{$\; \buildrel < \over \sim \;$}
\def\lsim{\lower.5ex\hbox{\ltsima}}
\def\gtsima{$\; \buildrel > \over \sim \;$}
\def\gsim{\lower.5ex\hbox{\gtsima}}
\def\mdot {\dot M}
\newcommand{\be}{\begin{equation}}
\newcommand{\en}{\end{equation}}
\newcommand{\ergs}{\rm \ erg \; s^{-1}}

\def\gs   {\rm \ g  \, s^{-1}}

\def\cmdue {\rm \ cm^{-2}}

\def\msole {~M_{\odot}}

\begin{document}

\received{~~} \accepted{~~}
\journalid{}{}
\articleid{}{}

\title{The quiescent X--ray emission of three transient X--ray pulsars}
\shortauthors{Campana et al.}

\author{S.~Campana\altaffilmark{1}, L. Stella\altaffilmark{2},
G.L. Israel\altaffilmark{2}, A. Moretti\altaffilmark{1},
A.N. Parmar\altaffilmark{3}, M. Orlandini\altaffilmark{4}}

\altaffiltext{1}{INAF--Osservatorio Astronomico di Brera, Via Bianchi 46, I-23807
Merate (Lc), Italy}

\altaffiltext{2}{INAF--Osservatorio Astronomico di Roma,
Via Frascati 33, I-00040 Monteporzio Catone (Roma), Italy}

\altaffiltext{3}{Astrophysics Division, Space Science Department of ESA,
ESTEC, P.O. Box 299, 2200 AG Noordwijk, The Netherlands}

\altaffiltext{4}{Istituto Tecnologie e Studio Radiazioni Extraterrestri, C.N.R.,
Via Gobetti 101, I--40129 Bologna, Italy}

\email{campana@merate.mi.astro.it}

\begin{abstract}
We report on BeppoSAX and Chandra observations of three Hard X--Ray Transients
in quiescence containing fast spinning ($P<5$ s) neutron stars: A~0538--66, 
4U~0115+63 and V~0332+53. These observations allowed us to study these
transients at the faintest flux levels thus far. Spectra are 
remarkably different from the ones obtained at luminosities a factor $>10$
higher, testifying that the quiescent emission mechanism is
different. Pulsations were not detected in any of the sources, indicating that
accretion of matter down to the neutron star surface has ceased. 
We conclude that the quiescent emission of the three X--ray transients likely
originates from accretion onto the magnetospheric boundary in the propeller
regime and/or from deep crustal heating resulting from pycnonuclear reactions
during the outbursts.
\end{abstract}


\section{Introduction}

Young magnetic neutron stars orbiting a Be star companion occasionally show
transient X--ray emission. 
Their spectra are relatively hard up to energies of tens of keV (power
law with photon indexes $\sim 1$) hence their name of Hard X--ray transients
(HXRTs; White, Kaluzienski \& Swank 1984). 
Be stars are characterised by equatorial mass loss episodes likely originating
from their high (nearly break up) rotational velocities. 
When the neutron star along
its (eccentric) orbit enters this equatorial disk an X--ray outburst episode is
observed (Stella et al. 1986; Bildsten et al. 1997). 
Part of the material outflowing from the Be star is
captured by the gravitational field of the neutron star, accretion onto
its magnetic polar caps takes place, generating an intense pulsed X--ray flux.
The spin periods of HXRTs range from 69 ms (A 0538--66; Skinner et al.
1982) to 25 min (RX J0146.9+6121; Mereghetti, Stella \& De Nile 1993). 
Magnetic field strengths, when accurately inferred from 
cyclotron line features, lie in the range of $B\sim (1-10)\times 
10^{12}$ G (Dal Fiume et al. 2000).

The powering mechanism and the accretion regime that pertains to the quiescent 
state of these sources is still uncertain. A change in the accretion  
regime should take place at lower luminosities in the case of 
fast spinning neutron stars as their magnetosphere expands beyond the
corotation radius, therefore halting the infalling material at the
magnetospheric boundary (i.e. the propeller regime). 
A contribution from matter leaking through the centrifugal barrier
may still be present (Campana et al. 2001). A further emission mechanism is
represented by cooling of the neutron star surface (deep crustal heating) 
due to pycnonuclear reactions occurring during outbursts (Brown, Bildsten \&
Rutledge 1998).

In this paper we present the first detailed investigation of the quiescent
state of three of the fastest spinning (accreting) neutron stars
in HXRTs. A 0538--66 and 4U 0115+63 were detected in our BeppoSAX
observations while V 0332+53 remained undetected. A Chandra pointing 
revealed also the latter source (Section 2).  
We discuss these results in the light of the regimes experienced by a neutron
star subject to a range of matter inflow. By using the neutron star parameters 
deduced from observations in outburst, we infer that these sources are likely 
in the propeller regime (Section 3). 

\section{X--ray observations}

We analysed the data from the two imaging instruments on board the BeppoSAX 
satellite: the Low Energy Concentrator Spectrometer (LECS; 0.1--10 keV, 
Parmar et al. 1997) and the Medium Energy  Concentrator Spectrometer 
(MECS; 1.6--10.5 keV, Boella et al. 1997). Non-imaging instruments (when active) 
provided only upper limits.
Only two of the three MECS units were operating at the time of the observations.
LECS data were collected only during satellite night-time resulting in shorter 
exposure times. For a summary of the observations see Table 1.

\begin{table*}
\label{table1}
\begin{center}
\caption{Summary of observations.}
\begin{tabular}{cccc}
\tableline
Source     & Start--Stop    & LECS          & MECS         \\ 
           & time           & Exp. time (s) & Exp. time (s)\\ 
\tableline
A 0538--66 &Sep. 10--11 1999& 12706         & 38540        \\ 
4U 0115+63 &Aug. 13--16 2000& 44295         & 85547        \\ 
V 0332+53  &Aug. 14--15 1999& 17748         & 39218        \\ 
\tableline
V 0332+53  &Jan. 04--04 2001& Chandra       &\  5146       \\
\tableline
\end{tabular}
\end{center}
\end{table*}

Products were extracted using the FTOOLS package (v. 5.0.1). LECS and MECS
events were extracted from a circle of $4'$ radius. The background was
subtracted using spectra from blank sky files at the same detector coordinates
(after checking that the background of the observation is comparable).   
For the spectral analysis we used the XSPEC software (v. 11.1.0) and the 
public response matrices available in January 2000. During the spectral fits 
a variable normalisation factor was included to
account for the mismatch in the absolute flux calibration of the BeppoSAX
instruments (Fiore, Guainazzi \& Grandi 1999).

\subsection{A 0538--66}

The X--ray transient A 0538--66 in the Large Magellanic Cloud contains a 69 ms 
accreting neutron star.
Pulsations were detected only once during a strong outburst in the late
seventies with peak luminosities of $\sim 10^{39}\ergs$  (2--17 keV;
Skinner et al. 1982; Ponman, Skinner \& Bedford 1984). An upper limit on the
magnetic field of $B\sim 10^{11}$ G  was inferred through the presence of a
pulsed signal as the source emitted a luminosity of $\sim 10^{39}\ergs$, under
the (very likely) hypothesis that matter, unimpeded by the centrifugal
barrier, accreted onto the neutron star surface at that time (Skinner et
al. 1982). 
A 5-yr optical light curve obtained as a by-product of the MACHO project
revealed a long-term modulation at $\sim 421$ d and confirmed the shorter term 
modulation at the orbital period of $16.651$ d (Alcock et al. 2001).
During the ROSAT all-sky survey two weaker outbursts from A 0538--66 were
detected, with peak luminosities of $\sim 4$ and $\sim
2\times10^{37}\ergs$ in the 0.1--2.4 keV range (Mavromatakis \& Haberl 1993). 
This source was also observed a few times in its quiescent state.
ASCA detected  A~0538--66 in a 40 ks observation close to periastron
at a level of $\sim 2\times 10^{36}\ergs$ (Corbet et al. 1997).
The spectrum could be described by the sum of a power law
(with photon index $\Gamma \sim 2$) plus a soft component (which could be fit
with a black body or a bremsstrahlung). An iron line had 
also to be included in order to obtain a good fit to the data.
Campana (1997) analysed ROSAT PSPC serendipitous pointings in which
A~0538--66 was (usually) detected at a level of $3-10\times 10^{34}\ergs$
(0.1--2.4 keV). 

The BeppoSAX observation of A 0538--66 was carried out on 1999 Sep. 10--11 and
covered an orbital phase interval of 0.92--0.98 according to the ephemerides
in Skinner et al. (1982) and Alcock et al. (2001). The phase of the 421 d
periodicity discovered by Alcock et al. (2001) was approximately 0.77. 
Outbursts are observed in only a limited phase range of this cycle. This
cycle is quasi-periodic and our observation was carried out during the
non-outbursting part of the cycle.

The source was detected in the LECS and MECS data with a net count rate
of $2\times 10^{-3}$ ct s$^{-1}$ each. The source contributed for $27\%$ and
$38\%$ of the total counts in the LECS and MECS $4'$ extraction circles.
The LECS and MECS spectra had a relatively poor
statistics, with 97 and 456 source counts, respectively. 
We rebinned the LECS and MECS spectra in order to have 40 and 50 counts 
per bin, respectively. A spectral analysis was carried out in the 0.1--8 keV
energy range for the LECS and 1.6--7 keV for the MECS. Single component models 
provided a good description of the data.
A power law model with a photon index $\Gamma=2.1\pm0.6$ (errors are
at $90\%$ confidence level for one parameter of interest; this is 
adopted throughout the text unless otherwise specified) and a column density
of $N_H<1.1\times 10^{22}\cmdue$ provided a reduced $\chi^2_{\rm red}=1.3$
for 5 degrees of freedom (d.o.f.; see Fig. \ref{fig1}).
A bremsstrahlung model could fit the data equally well with an equivalent
temperature $T_{\rm br}=5.5^{+9.6}_{-2.9}$ keV and
$N_H<5.8\times10^{21}\cmdue$, resulting in a $\chi^2_{\rm red}=1.1$ (even if
the nominal column density is consistent with zero).
A black body model with an equivalent temperature of 
$T_{\rm bb}=0.8\pm0.2$ keV and a null column density 
($N_H<4.4\times 10^{21}\cmdue$) gave instead a $\chi^2_{\rm red}=2.4$.
The equivalent black body radius is $R_{\rm bb}=1.4^{+0.7}_{-0.5}$ km. 
For all the models the 0.1--10 keV unabsorbed luminosity ranged between 
$1-3\times 10^{35}\ergs$ for a distance of 50 kpc.

\begin{inlinefigure}
\bigskip
\centerline{\includegraphics[width=1.2\linewidth]{0538_quie.ps}}
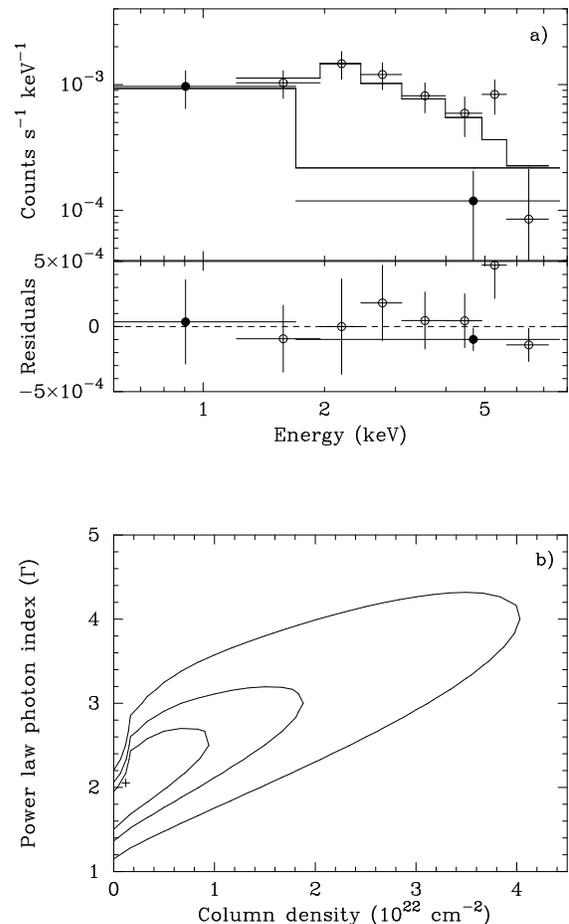
\caption{Panel $a$): X--ray spectrum of A 0538--66 as observed by BeppoSAX. 
Filled and open circles refers to LECS and MECS data, respectively. The 
overlaid model is a power law with interstellar absorption (see text). 
Residuals are shown in the lower panel. 
Panel $b$): contour plot of the power law photon index versus the column
density. Contours are at a level of 1, 2 and 3 $\sigma$.}
\label{fig1}
\bigskip
\end{inlinefigure}

The 456 MECS photons were used for the timing analysis.
Photon arrival times were first corrected to the bary\-center of the solar system.
We then searched for pulsations at the neutron star spin period (69 ms;
Skinner et al. 1982). To decide the period range over which the search should be
carried out we estimated the variation of the period since the last
observation in which pulsations were detected. The value of the period
derivative measured during the 1979 outburst $\dot{P}=5\times 10^{-10}$ s
s$^{-1}$ (Skinner et al. 1982) was too large to derive from accretion torques
alone and was likely dominated by Doppler shifts. We adopt here as a scale
value the $\dot{P}$ measured for 4U 0115+63 (see below).  
Due to the short spin period and long time elapsed since the last
outburst (1982) the frequency range spanned with the assumed maximum $\dot{P}$
is fairly large, encompassing the range 64--74 ms.
No periodicities were found due to poor statistics.

The BeppoSAX observation detected A 0538--66 at its
lowest flux level in the 0.1--10 keV range (a factor of $\sim 50$ less 
than in the ASCA observation; Corbet et al. 1997).
The BeppoSAX spectrum was significantly different from that measured by ASCA. 
Assuming the ASCA spectral parameters and leaving the 
normalisation free, a $\chi^2_{\rm red}=7.0$ was obtained. 
The main discrepancy was the absence of the soft component, which was 
clearly required by the ASCA data but unnecessary in the BeppoSAX spectrum.
Setting the black body normalisation to zero, the $\chi^2_{\rm red}$ 
decreased to 1.4, i.e. the best fit derived above. 
The other feature found by Corbet et al. (1997), i.e. a 
highly ionised iron line (with equivalent width $EW=450$ eV), was not
detected in our spectrum with an upper limit of $EW \lsim 400$ eV.

\subsection{4U 0115+63}

When in outburst, the X--ray flux of 4U~0115+63 is modulated at the spin period 
of 3.6~s (Cominsky et al. 1978). Pulse arrival time analysis yielded 
an orbital solution with a period of 24.3 d and an eccentricity of 0.34 
(Rappaport et al. 1978).
Observations with {\it Ginga} confirmed earlier results (Wheaton et al.
1979; White, Swank \& Holt 1983) of cyclotron line features
characteristic of a high magnetic field $B\sim 10^{12}$ G (Nagase 
1991). BeppoSAX observations led to the detection of four cyclotron line
harmonics, corresponding to a magnetic field of $B=1.3\times 10^{12}$ G at the
neutron star surface (after correction for the gravitational redshift;
Santangelo et al. 1999). 

Tight upper limits on the quiescent emission of this source were
derived from Einstein, EXOSAT and ROSAT data (Campana 1996).
The column density to the source is high ($\sim 10^{22}\cmdue$) and limits 
the possibility of detecting and studying a possible soft X--ray component. 

During a BeppoSAX observation close to periastron we revealed a huge
luminosity increase (factor $\gsim 250$) in less than 15 hr. This was
interpreted in terms of the opening of the centrifugal barrier in the  
transition regime between propeller and accretion regimes, thus providing the
best evidence to date for the onset of the propeller mechanism (Campana et
al. 2001). 

A further BeppoSAX observation was carried out on 2000 Aug. 13--16 (a short
account of the results was reported in Campana et al. 2001).
The observation covered the orbital phase range 0.42--0.51 (adopting the
ephemerides of Bildsten et al. 1997). The source was not detected in the LECS
(likely due to high absorption) but was seen in the MECS data.
The $3\,\sigma$ upper limit on the LECS count rate was $2\times 10^{-3}$ ct 
s$^{-1}$. 
The net source count rate in the MECS was $10^{-3}$ ct s$^{-1}$ and it
comprised only $\sim 15\%$ of the total flux in the $4'$ extraction 
region and in the 1.4--9.5 keV energy band
due to the large background. We rebinned the spectrum in order to have 150
counts per channel, resulting in 6 energy bins. The spectrum was very poor and
could be fit by a variety of single-component models. We fixed the column
density to the best fit value found during the observation at periastron,
i.e. $N_H=1.74\times 10^{22}\cmdue$ (Campana et  al. 2001). In the case of a
power law model we obtained a photon index $\Gamma=2.6^{+2.1}_{-1.3}$
($\chi^2_{\rm red}=0.2$ for 1 d.o.f.; see Fig. \ref{fig2}).
Bremsstrahlung and black body models both provided good fits but with a column 
density well below the value observed in outburst. In particular, 
for the bremsstrahlung model we obtained $T_{\rm br}=2.8^{+\infty}_{-2.0}$ keV 
($\chi^2_{\rm red}=0.2$ for 1 d.o.f.) and $T_{\rm bb}=0.7^{+0.7}_{-0.3}$ keV 
($\chi^2_{\rm red}=0.5$ for 1 d.o.f.). 
All models provided an unabsorbed 0.5--10 keV luminosity of $0.8-2 \times
10^{33}\ergs$ for a distance of 8 kpc (Negueruela \& Okazaki 2001).

\begin{inlinefigure}
\centerline{\includegraphics[width=1.2\linewidth]{0115_quie.ps}}
\vskip -3truecm
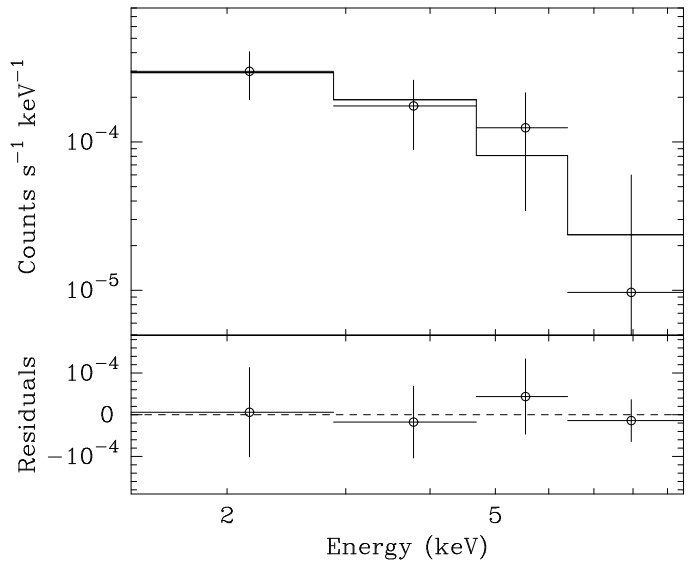
\caption{X--ray spectrum of 4U 0115+63 as observed by BeppoSAX MECS. The 
overlaid model is a power law with interstellar absorption (see text). 
Residuals are shown in the lower panel.} 
\label{fig2}
\end{inlinefigure}

 From the same extraction region we selected 690 events for the timing analysis.
The photon arrival times were corrected to the barycenter of the solar system.
Pulsations at the neutron star spin period were searched for. We consider a 
period interval of 3.616--3.617 s derived from the spin period detected
in Aug. 1999 (Campana et al. 2001) and a maximum spin-up or spin-down
rate of $8\times 10^{-12}$ s s$^{-1}$ (Bildsten et al. 1997).
The search over 13 independent Fourier frequencies provided an upper limit
on the pulsed fraction of $30\%$ ($3\,\sigma$). 
\begin{table*}
\label{table2}
\begin{center}
\caption{Relevant luminosities of HXRTs of our sample.}
\begin{tabular}{c|cccccc}
\tableline
Source     &Spin period& $B$$^*$          &$\log L_{\rm min}(R)$&Centr. gap &$\log L_{\rm min}(r_{\rm cor})$& $\log L_{\rm obs}$\\ 
           &  (s)      &($10^{12}$ G)     &  (erg s$^{-1}$)     &($\Delta$) &(erg s$^{-1}$)                 & (erg s$^{-1}$)   \\
\tableline
A 0538--66 &  0.07  & $\sim 0.1$          &$<$ 37.9--38.9 &\  28         & $<$ 36.4--37.5& 35.3 \\
4U 0115+63 &  3.62  &    1.3              &\ \ 35.5--36.5 &  400         &\ \  32.9--33.9& 33.0 \\
V 0332+53  &  4.37  &    3.5              &\ \ 36.1--37.2 &  450         &\ \  33.5--34.5& 32.7 \\
\tableline
\end{tabular}
\end{center}

\tablenotetext{}{Luminosity ranges have been computed for the limiting cases 
$\xi=0.5$ and $\xi=1$.}

\tablenotetext{*}{Magnetic field was  estimated from cyclotron line
features. In the case of A 0538--66 it was estimated indirectly from the 
detection of pulsations during a very luminous outburst (see text) and
therefore adopt a magnetic field value consistent with the estimate of Skinner
et al. (1982) and our definitiion of $B$, i.e. a factor of two higher.}

\end{table*}

\subsection{V 0332+53: BeppoSAX observation}

EXOSAT observed three outbursts from V 0332+53 between November 1983 and
January 1984, leading to the discovery of the 4.4 s spin periodic and a sudden
decrease of luminosity at the end of $\sim 1$ month long recurrent
outbursts. The latter result was interpreted in terms of the onset of the
centrifugal barrier (Stella et al. 1985, 1986). An upper limit of $\sim
5\times 10^{33}\ergs$ to the source quiescent emission (1--15 keV) was derived
on that occasion with the EXOSAT ME.  
Doppler shifts in pulse arrival times indicate that the pulsar is in orbit
around a Be star with period of 34.3 d and eccentricity 0.3 (Stella et al. 1985).
Observations during a subsequent outburst with {\it Ginga} led to the
discovery of a cyclotron line feature corresponding to a $\sim 3\times
10^{12}$ G magnetic field (Makashima et al. 1990). 

The BeppoSAX observation took place on 1999 Aug. 14--15.  
The current orbital ephemerides are not accurate enough to assess the orbital
phase interval of this observation.
V~0332+53 was not detected in the LECS and MECS instruments (see below). 
A source unrelated to V~0332+53 was revealed in the MECS data $\sim 6'$
off-axis. 
The LECS and MECS exposures provided a $3\,\sigma$ upper limit on the V 0332+53 
count rate of $2\times 10^{-3}$ ct s$^{-1}$. This translates into an unabsorbed 
luminosity limit of $\lsim 3\times 10^{33}\ergs$ (0.1--10 keV), assuming a 
power law spectrum ($\Gamma=2$), a column density of $N_H=1\times 
10^{22}\cmdue$ (consistent with the value observed in outburst;
Makashima et al. 1990), and a distance of 7 kpc (Negueruela et al. 1999).
A harder power law with $\Gamma=1$ results in a factor of 4 higher upper limit.

\subsubsection{V 0332+53: Chandra observation}

V 0332+53 was observed by Chandra on 2001 Jan. 04 with the ACIS-S instrument 
for 5 ks. Despite the short exposure time the superb angular
resolution of the Chandra telescope afforded a much fainter limiting 
flux. Our Brera Multiscale Wavelet detection algorithm tailored for Chandra
(BMW-Chandra, Moretti et al. 2002) was used to detect the source. V 0332+53
was detected at the very low level of $4.4\times 10^{-3}$ ct s$^{-1}$ (22
counts). Assuming a power law spectrum with $\Gamma=2$ and a column density of
$10^{22}\cmdue$, this rate converts to a 0.5--10 keV unabsorbed flux of $\sim
9\times 10^{-14}\ergs\cmdue$. The corresponding luminosity was $\sim 5\times
10^{32}\ergs$. A harder power law with $\Gamma=1$ resulted in a factor of 3
larger luminosity. This is the faintest luminosity yet observed from a HXRT in
quiescence. Its value is comparable to the level observed in low mass
transients containing an old, weakly magnetic, fast spinning neutron star
(e.g. Campana et al. 1998). The softness ratio (0.5--2 keV)/(2--6 keV) was
$2.6\pm1.4$. Assuming the same column density as in outburst we constrain 
the power law photon index within the 2--3.5 range ($68\%$).

\section{Discussion}

We observed a sample of fast spinning neutron stars in HXRTs during
quiescence with BeppoSAX and Chandra.
The quiescent luminosities observed in the fast HXRTs of our sample
are very low (especially in the case of V 0332+53, see Table 2) 
and one can
ask if the inflowing matter can reach the neutron star surface. If accretion
onto the neutron star surface took place in quiescence, then
the mass inflow rate has to decrease by a large factor (up to $10^6$ in the
case of V 0332+53) from outburst to quiescence, posing severe limitations to
the Be wind and disk characteristics.
A different way out is represented by the limited efficiency of the
accretion process because matter is halted at the neutron star
magnetosphere ($r_{\rm m}$) when the magnetic field lines rotate locally at  
super-Keplerian speed. This process is often referred to as the centrifugal
barrier or propeller mechanism (Illarionov \& Sunyaev 1975; Stella et al. 1986).
These systems have well known spin periods and magnetic field strengths thus
allowing us to estimate the luminosity at which the centrifugal
barrier starts operating.  
Using simple spherical accretion theory(which also provides 
a good approximation in the case of disk accretion, e.g. Wang 1995, 1996), 
one can work out the limiting mass inflow rate $\mdot_{\rm lim}$ 
and in turn the limiting accretion luminosity for the onset of the propeller:
\begin{eqnarray}
L_{\rm lim}(R)&=&G\,M\,\mdot_{\rm lim}/R \\ \nonumber &\simeq& 3.9\times 10^{37}
\,\xi^{7/2}\,B_{12}^2\,P_0^{-7/3}\,M_{1.4}^{-2/3}\,R_6^{5}\ergs 
\end{eqnarray}
(where the neutron star magnetic field, spin period, mass and radius are
scaled as $B=B_{12}\,10^{12}$ G\footnote{The magnetic field is obtained from
the magnetic dipole moment $\mu=B\,R^3/2$.}, $P=P_0$ 1 s,
$M=M_{1.4}\,1.4\msole$ and $R=R_6\,10^6$ cm, respectively, e.g. Stella et
al. 1986). $\mdot$ indicates the mass accretion rate and $G$ is the
gravitational constant. The factor $\xi$ accounts for the deviations of
$r_{\rm m}$ as computed in spherical symmetry from the case of an accretion
disk. In general $\xi$ is in the range 0.5--1.5 (here we use $\xi=1$).
Values in the range of $\sim 10^{34}-10^{37}\ergs$ are derived for the onset 
of the centrifugal inhibition in the fast HXRTs of our sample (see below and 
Table 2).

For lower mass inflow rates than those in Eq. 1, the great majority of
accreting matter can no longer reach the neutron star surface 
and a sharp drop off of the accretion luminosity is expected. 
The corresponding luminosity jump mainly depends on the spin
period of the neutron star 
\be
\Delta=\bigl({{G\,M\,P^2}\over {4\,\pi^2\,R^3}}\bigr)^{1/3} = 170\, 
M_{1.4}^{1/3}\,R_6^{-1}\,P_0^{2/3}
\en
(e.g. Corbet 1996; Campana \& Stella 2000) and is a factor of 
30--500 for the HXRTs in our sample (see Table 2). Therefore, the
maximum accretion luminosity that can be emitted in the propeller regime is 
\begin{eqnarray}
L_{\rm min}(r_{\rm cor})&=&L_{\rm min}(R)/\Delta=G\,M\,
\mdot_{\rm lim}/r_{\rm cor}= \\ \nonumber
&=&2.4\times 10^{35}\,\xi^{7/2}\,B_{12}^2\,P_0^{-3}\,M_{1.4}^{-1}\,R_6^{6}\ergs 
\end{eqnarray}
Clearly these luminosities are all bolometric. While in the case of accretion
onto the neutron star surface most of the emission goes into X--rays,  
in the propeller regime this is not clear and these numbers should be referred
as upper limits. Moreover, the physics of the propeller regime is poorly
understood and a fraction of matter may still leak through the barrier.
 
As can be noted from Table 2, the observed X--ray luminosities are
all below the threshold for the onset of the centrifugal barrier and below the 
maximum expected luminosity in the propeller regime (this is true even for  
$\xi=0.5$). This testifies that the HXRTs in our sample are all likely detected
in the propeller regime and the observed luminosity derives from the
mass inflow releasing its gravitational energy down to the magnetospheric radius.

The luminosity level pertaining to quiescence in the propeller regime
clearly depends on the unknown quiescent mass inflow. An additional and
independent luminosity can derive from the cooling of
the neutron star made hot during the events of intense accretion: the
inner crust compressed by the loaded material becomes the site of pycnonuclear
reactions that may deposit enough heat into the core (Brown, Bildsten \&
Rutledge 1998; Colpi et al. 2001; see also Campana et al. 1998). 
In the last few years V 0332+53 and A 0538--66 did not show any outburst
activity and therefore it is hard to estimate a mean accretion rate. This is
instead possible for 4U 0115+63 which showed two strong outbursts and a number 
of small outbursts during the RXTE lifetime. 
Based on the observed outbursts one can derive a time-average rate of
$\sim 4\times10^{15}\gs$, resulting in a deep crustal heating luminosity of
$4\times10^{33}\ergs$. This value has to be compared with the inferred black
body luminosity of $\sim 10^{33}\ergs$, which is a factor $\sim 4$ lower, this 
luminosity however could be hidden in the lower energy part of the spectrum.
Similar luminosity levels (if not lower) apply to V 0332+53 and A 0538--66. 
Being the quiescent luminosity A 0538--66 $\sim 5\times 10^{35}\ergs$ it
cannot be supported by this emission mechanism only.

\end{document}